\documentclass[aps,10pt,showpacs,twocolumn]{revtex4}
\usepackage{amsmath, amssymb, lipsum}
\usepackage{graphics, graphicx, epsfig, color, subfigure}
\usepackage{bbold}
\begin{document}

\title{Axial anomaly effects in finite isospin $\chi$PT in a magnetic field}

\date{\today}

\author{Prabal Adhikari}
\email{adhika1@stolaf.edu}
%\email{prabal.adhikari@gmail.com}
\affiliation{St. Olaf College, Physics Department, 1520 St. Olaf Avenue, Northfield, MN 55057}


\begin{abstract}
%In this paper, we consider the phase diagram of low energy, finite-isospin QCD in the context of leading order chiral perturbation theory in a uniform magnetic field. In particular, we consider the Schwinger limit, where photons cannot back-react to the external magnetic field. We prove rigorously that only neutral pions condensates can form in the Schwinger limit. We show that $\pi^{0}$ domain walls are an example of such a state and find the magnetic fields above which they become metastable. Furthermore, we incorporate the effects of the axial anomaly through the Wess-Zumino-Witten term and show that above a critical baryon chemical potential, which was first calculated in Ref.~\cite{Son4}, the phase diagram of finite-isospin-chiral perturbation theory away from the Schwinger limit, which is effectively similar to that of a type-II superconductor, is modified by the presence of $\pi^{0}$ domain walls that becomes stable at the expense of the normal state above the critical baryon chemical potential.
In this paper, we consider finite isospin chiral perturbation theory including the effects of the axial anomaly (through the Wess-Zumino-Witten term) in a strong magnetic field. We firstly prove that in a strong external magnetic field ($H_{\rm ext}$) or more precisely the Schwinger limit, where photon back-reactions are suppressed, only neutral pions can condense and the condensation of charged pions is forbidden. Secondly, we find that the $\pi^{0}$ domain wall is an example of a phase that can exist in a strong magnetic field and suggest the existence of a new phase transition line from the normal vacuum state to the $\pi^{0}$ domain wall state. This phase transition exists for non-zero pion masses if the baryon chemical potential exceeds a critical value $16\pi f_{\pi}^{2}m_{\pi}/eH_{\rm ext}$. The phase transition line persists away from the Schwinger limit when the photons can back-react to the external magnetic field.
\end{abstract}

\maketitle
\section{Introduction}
The fundamental degrees of freedom of Quantum Chromodynamics (QCD) are quarks, which couple to magnetic fields through their electromagnetic charge. As a result, QCD exhibits some interesting phase structure in the presence of magnetic fields~\cite{Chernodub}. RHIC collisions between nuclei generate large magnetic fields; here chiral magnetic effect is expected to play a significant role~\cite{Naoki}. Similarly, the color-flavor-locked phases~\cite{Gorbar,Son5}, which can exist (at least theoretically) at asymptotically large baryon densities may be found in neutrons stars and also possibly in magnetars, where the magnetic fields are to the order of $10^{18}\ G$~\cite{Harding}. Furthermore, the quark-gluon plasma of the early universe may have sustained large magnetic fields. As such there is a lot of interest in the properties of QCD vacuum and other QCD phases in the presence of strong magnetic fields.

QCD in a strong magnetic field is a topic of great contemporary interest and has been studied in the Schwinger limit quite extensively~\cite{Schwinger}. In this limit, the quantity $qH$ is held fixed (with $q$ being a relevant electromagnetic charge and $H$ being the external field) while $H\rightarrow\infty$ as $q\rightarrow 0$. This limit was first considered in the context of vacuum pair production in QED by Schwinger~\cite{Schwinger} but since has been used to study the low-energy QCD vacuum in the context of chiral perturbation theory, primarily for mesons. For instance, the chiral condensate of the low-energy QCD vacuum has been studied through chiral perturbation theory using the Schwinger limit for zero and non-zero pion masses up to next-to-leading order and the result has been generalized to the finite temperature case~\cite{Shushpanov1,Shushpanov2,Werbos1,Werbos2}. At large magnetic fields, QCD undergoes dimensional reduction and quarks acquire masses of $\mathcal{O}(\sqrt{qH})$, where $q$ is the quark electromagnetic charge and $H$ the external field. This explains the origin of magnetic catalysis due to the ``Cooper effect", which means that pairing is more effective in one spatial dimension compared to three~\cite{Shovkovy}.

Recently, there has been interest in finite isospin QCD, where the QCD vacuum has been shown to exhibit superconducting behavior through the condensation of charged pions~\cite{Adhikari, Endrodi}. This has been shown in the context of chiral perturbation theory and lattice QCD, even though the lattice seems to have a sign problem in the presence of both an isospin chemical potential and a magnetic field due to the charge difference between the up and down quarks. 

The study of Ref.~\cite{Adhikari} was performed in chiral perturbation theory at finite isospin chemical potentials and a uniform magnetic field, with the charged pions coupled to dynamical photons. It was shown that for realistic pion masses, the system behaves as a type-II superconductor, with a uniform superconducting phase at low magnetic fields and a phase transition to a normal phase with increasing magnetic fields through an intermediate phase of topological vortices~\cite{Gorkov, Landau, Abrikosov, Tinkham, Fetter}. The study was performed at leading order in chiral perturbation theory and is valid as long as the relevant physical parameters including the pion mass ($m_{\pi}$), the isospin chemical potential ($\mu_{I}$) and the magnetic field ($H$) satisfy the condition that 
\begin{equation}
m_{\pi},\ \mu_{I},\ \sqrt{eH}\ll \Lambda_{\rm Had}\ ,
\end{equation}
where $\Lambda_{\rm Had}$ is the typical hadronic scale which of the order $4\pi f_{\pi}$, with $f_{\pi}$ being the pion decay constant.

In this paper, we consider chiral perturbation in a background magnetic field in the Schwinger limit. In particular, this means that the charged pions are coupled to the classical external magnetic fields at all orders. However, unlike the study of Ref~\cite{Adhikari}, we will assume that there are no dynamical photons that can couple to the charged pions. It is well-understood that superconductivity arises due to spontaneous symmetry breaking, whereby photons acquire mass. From the perspective of the relevant Lagrangian, a term proportional to $\vec{A}^{2}$ is introduced - it becomes impossibly expensive to sustain a magnetic field. Obviously, in the absence of photons and hence a back-reaction from the photons to a strong, external magnetic field, (i.e. the Schwinger limit), it is natural to ask what possible phases \textit{can} be sustained in finite isospin chiral perturbation theory. We will show in this paper that unlike the result of Ref.~\cite{Adhikari}, where the charged pions condensed, that in the Schwinger limit (with no photons), that charged pion condensation is forbidden and that only the neutral pions can condense.

We will also consider a particular example of the pion condensed phase, namely $\pi^{0}$ domain walls that was first considered in Ref.~\cite{Son4}. It was shown that this phase becomes more stable than the normal phase of chiral perturbation theory above a critical magnetic field. We will show a new phase transition line emerges in finite isospin chiral perturbation theory if the system is above the critical baryon chemical potential of Ref.~\cite{Son4} assuming that the $\pi^{0}$ domain wall is the lowest energy pion condensed phase that can exist. It is important to note that this assumption while plausible is strictly speaking unproven; the plausibility is motivated by the absence of other possible neutral pion phases that satisfy the equations of motion of finite isospin chiral perturbation theory.

The paper is organized as follows: in the first section, we summarize the chiral perturbation theory Lagrangian at finite isospin and a uniform magnetic field using the same notation as in Ref.~\cite{Adhikari}. In the following section, we will prove a no-go theorem that charged pions cannot condense in the Schwinger limit (with no back-reacting photon fields). The result seems plausible based on the fact that the mechanism that gives rise to photon masses through symmetry breaking (either spontaneous or explicit) is absent in the Schwinger limit. In the third section, we will consider how the existence of a $\pi^{0}$ domain state modifies the finite isospin low-energy QCD phase diagram in a magnetic field above the critical baryon chemical potential of Ref.~\cite{Son4}. In particular we will show a new phase transition from the normal state to a $\pi^{0}$ domain wall state, which is different from the result of Ref.~\cite{Son4} but reduces to the result of Ref.~\cite{Son4} at zero isospin chemical potential.

\section{Lagrangian of Finite Isospin Chiral Perturbation Theory in a Magnetic Field}
We begin with the leading order chiral perturbation theory Lagrangian at finite isospin and a magnetic field. The Lagrangian at finite isospin (but zero external magnetic field) was first considered in Refs.~\cite{Son1, Son2}, where it was shown that for isospin chemical potentials greater than or equal to the pion mass, a superfluid phase of condensed, charged pions becomes energetically favorable. It was shown that the charged superfluid in the presence of an external magnetic field with the pions coupled to back-reacting photons exhibit type-II superconductivity in Ref.~\cite{Adhikari}. The Lagrangian is as follows:
\begin{equation}
\label{Lagrangian}
\mathcal{L}_{\rm eff}=f_{\pi}^{2}\ \textrm{Tr} (D_{\mu}\Sigma^{\dagger}D^{\mu}\Sigma)+m_{\pi}^{2}f_{\pi}^{2}\textrm{Tr}(\Sigma+\Sigma^{\dagger}) \ ,
\end{equation}
where $f_{\pi}$ is the pion decay constant $m_{\pi}$ is the pion mass and $\Sigma$ are the $SU(2)$ matrices, which we parametrize as in Ref.~\cite{Adhikari}:
\begin{equation}
\label{Sigma}
\Sigma=\frac{1}{f_{\pi}}\left( \sigma\mathbb{1}+i\pi_{x}\tau_{1}+i\pi_{y}\tau_{2}+i\pi_{z}\tau_{3} \right )\ ,
\end{equation}
where the $\sigma$ and $\pi_{i}$ fields are defined as:
\begin{equation}
\label{Sigma}
\begin{split}
\sigma&=f_{\pi}\cos\psi\cos\theta\\
\pi_{x}&=f_{\pi}\cos\psi\sin\theta\cos\alpha\\
\pi_{y}&=f_{\pi}\cos\psi\sin\theta\sin\alpha\\
\pi_{z}&=f_{\pi}\sin\psi\ .
\end{split}
\end{equation}
Note that this parametrization guarantees that $\det\Sigma=1$ and $\Sigma^{\dagger}\Sigma=1$.

The gauge derivative in the Lagrangian incorporates both the finite isospin chemical potential and the finite isospin chemical potential. It is defined as follows:
\begin{equation}
D_{\mu}\Sigma=\partial_{\mu}\Sigma-i\ [\delta_{\mu 0}\mu_{\rm I},\ \Sigma]-ieA_{\mu}[Q,\ \Sigma]\ ,
\end{equation}
where isospin enters as a spatially-independent zeroth component of the gauge field, while the magnetic field enters through only the spatial components. Note that $e$ is the charge of a positive pion and the charge matrix for the up and down quarks is defined through the following relation:
\begin{equation}
Q=\frac{1}{6}\mathbb{1}+\frac{1}{2}\tau_{3}\ ,
\end{equation}
where $\mathbb{1}$ is a $2\times 2$ identity matrix and $\tau_{3}$ is the third Pauli matrix. Since the charge matrix $Q$ enters the Lagrangian through a commutator with the pion fields $\Sigma$, only the component of charge in the $\tau_{3}$ direction affects the Lagrangian.

We use the parametrization for $\Sigma$ defined in Eq.~(\ref{Sigma}) and couple the Lagrangian to the external magnetic field, which we will assume is uniform in space. In doing so, the effective Lagrangian of low-energy QCD at lowest order in chiral perturbation theory becomes:

\begin{equation}
\label{Leff}
\begin{split}
\mathcal{L}_{\rm eff}&=-\frac{f_{\pi}^{2}}{2}\left [ \cos^{2}\psi \{\sin^{2}\theta\left (\vec{\nabla}\alpha+e\vec{A} \right)^{2} + (\vec{\nabla}\theta)^{2}\} +(\vec{\nabla}\psi)^{2}\right ]\\&+m_{\pi}^{2}f_{\pi}^{2}(\cos\theta\cos\psi-1)+\frac{\mu_{I}^{2}f_{\pi}^{2}}{2}\sin^{2}\theta\cos^{2}\psi\\
&-\frac{1}{4}F_{ij}F^{ij}\ ,
\end{split}
\end{equation}
where we have assumed that the electromagnetic tensor only has spatial components, i.e. $F_{0i}=F_{i0}=0$. Note that we have included the kinetic contributions of the external field, which wasn't present in Eq.~(\ref{Lagrangian}).

\section{Phases in the Schwinger Limit}
In this section, we argue that the only possible phases that can exist at leading order in chiral perturbation theory in the Schwinger limit (i.e. no back-reacting photons) are phases where the charge pions remain uncondensed, in other words, we will prove that $\pi_{x}=\pi_{y}=0$ for all possible phases that can exist in the Schwinger limit. This seems somewhat reasonable: the mechanism that was responsible for superconductivity, namely photon fields acquiring mass through spontaneous symmetry breaking, is absent now since photons cannot react to the presence of a magnetic field. Then it becomes favorable for the system to remain in the ground state with $\theta=0\ ,\psi=0$ ($\sigma=f_{\pi},\ \pi_{x}=\pi_{y}=\pi_{z}=0$), even for chemical potentials larger than pion masses.

In order to prove that this is indeed the case, It is useful to begin with the equation of motion for the gauge fields $\vec{A}$, which can be easily derived using the effective Lagrangian of Eq.~(\ref{Leff}). The equation of motion for the gauge fields is
\begin{equation}
\label{EoMphoton}
-\partial_{k}F_{kl}=j_{l}\ ,
\end{equation}
where $F_{kl}$ refers to the spatial components of the electromagnetic tensor and $j^{l}$ refers to the components of the 3-current, which is as follows:
\begin{equation}
j_{l}=e \cos^{2}\psi\sin\theta\left (\partial_{l}\alpha+eA_{l} \right )\ .
\end{equation}
In the Schwinger limit (with no back-reactions), the photon fields $A$ must equal the external field $A_{\rm ext}$, which for a uniform magnetic field is of the following form:
\begin{equation}
A^{\rm ext}=\left (-b y H_{\rm ext},\ a x H_{\rm ext},\ 0\right )\ ,
\end{equation}
where $H_{\rm ext}$ is the external magnetic field with $a$ and $b$ satisfying the constraint $a+b=1$. In other words, we are working in the most general gauge configuration for a uniform magnetic field pointing in the positive z-direction. The no-back-reaction constraint on the electromagnetic tensor then is:
\begin{equation}
F_{kl}=\partial_{k}A_{l}^{\rm ext}-\partial_{l}A_{k}^{\rm ext}\ .
\end{equation}
Since the electromagnetic tensor must be spatially homogeneous, Eq.~(\ref{EoMphoton}) implies that the only way the equation of motion for photons is satisfied then is if the 3-current, $j^{l}$, vanishes. The constraint is satisfied only if at least one of the constraints given below is satisfied. Either
\begin{equation}
\begin{split}
\label{scenarios}
\textrm{A: } &\psi=\frac{\pi}{2} \textrm{ or}\\
\textrm{B: } &\theta=0 \textrm{ or}\\
\textrm{C: }  &\partial_{l}\alpha+eA^{\rm ext}_{l}=0.
\end{split}
\end{equation}
Scenario A corresponds to a chirally restored phase with the vacuum pointing entirely in the $\tau_{3}$-direction, i.e. $\pi_{z}=f_{\pi}$ and $\sigma=\pi_{x}=\pi_{y}=0$, a phase where the neutral pion condensate is at saturation. Scenario B corresponds to a phase with $\pi_{x}=\pi_{y}=0$, i.e. a phase which only allows the condensation of neutral pions (and cannot be chirally restored as we will see due to the exclusion of Scenario A). Finally scenario C is a constraint on the complex phase of one of the charge pions, either $\pi^{+}$ or $\pi^{-}$.

\subsection{Exclusion of Scenarios A and C}
Here we will argue that it is not possible to have any stable solutions for either scenario A or C.\\
\\
\textit{Scenario A:}
In order to rule out the chirally restored phase implied by scenario $A$, it is important to consider all possible phases for which the condition $\psi=\frac{\pi}{2}$ is satisfied. We begin with the equation of motion for $\psi$, which is as follows:
\begin{equation}
\begin{split}
&\vec{\nabla}^{2}\psi+\sin\psi\cos\psi\{ \sin^{2}\theta(\vec{\nabla}\alpha+e\vec{A})^{2}+(\vec{\nabla}\theta)^{2}\}\\
&-m_{\pi}^{2}\sin\psi\cos\theta-\mu_{I}^{2}\sin^{2}\theta\sin\psi\cos\psi=0 \ .
\end{split}
\end{equation}
It is straightforward to see from this equation that the only way in which the equation of motion is satisfied for scenario A away from the chiral limit is if $\theta=\frac{\pi}{2}$. However, it is easy to show that this phase is unstable. Since this phase is spatially homogeneous, we proceed by considering the effective potential, which can be deduced from the Lagrangian of Eq.~(\ref{Leff}) to be:
\begin{equation}
V_{\rm eff}=-m_{\pi}^{2}f_{\pi}^{2}\left (\cos\theta\cos\psi-1 \right )-\frac{\mu_{I}^{2}f_{\pi}^{2}}{2}\sin^{2}\theta\cos^{2}\psi\ .
\end{equation}
Note that $V_{\rm eff}$ has been normalized such that it is zero for $\theta=\psi=0$.

In order to proceed note that the first partial derivatives of the effective potential are
\begin{equation}
\label{firstderivative}
\begin{split}
\frac{\partial V_{\rm eff}}{\partial \theta}&=m_{\pi}^{2}f_{\pi}^{2}\sin\theta\cos\psi-\mu_{I}^{2}f_{\pi}^{2}\sin\theta\cos\theta\cos^{2}\psi\\
\frac{\partial V_{\rm eff}}{\partial \psi}&=m_{\pi}^{2}f_{\pi}^{2}\cos\theta\sin\psi+\mu_{I}^{2}f_{\pi}^{2}\sin^{2}\theta\cos\psi\sin\psi\ .
\end{split}
\end{equation}
Note that when $\theta=\psi=\frac{\pi}{2}$, $\frac{\partial V_{\rm eff}}{\partial\theta}=\frac{\partial V_{\rm eff}}{\partial\psi}=0$, as expected. However, using the second partial derivative test, the discriminant at $\theta=\psi=\frac{\pi}{2}$ assumes the following value:
\begin{equation}
\frac{\partial^{2}V_{\rm eff}}{\partial \theta^{2}}\frac{\partial^{2}V_{\rm eff}}{\partial \psi^{2}}-\left (\frac{\partial^{2}V_{\rm eff}}{\partial\theta\partial\psi} \right )^{2}=-f_{\pi}^{4}m_{\pi}^{4}\ ,
\end{equation}
which suggests that the solution is a saddle point assuming pion masses are non-zero. In the chiral limit, the second partial derivative test is inconclusive. However, it is easy to see from Eq.~(\ref{firstderivative}) that the point remains a saddle point even in the chiral limit.\\
\\
\noindent
 \textit{Scenario C:}
Next, we exclude scenario C because it is not possible to find an $\alpha$ that consistently satisfies the three constraints implied by Eq.~(\ref{scenarios}), scenario C. Explicitly the three constraints are:
\begin{equation}
\begin{split}
\label{scenarioC}
\partial_{x}\alpha&=eayH_{\rm ext}\\
\partial_{y}\alpha&=-ebxH_{\rm ext}\\
\partial_{z}\alpha&=0\ .
\end{split}
\end{equation}
Note that the last constraint above implies that $\alpha$ is independent of $z$. Then, solving the first constraint in the equation above, we get the following functional form for $\alpha$:
\begin{equation}
\label{alpha}
\alpha(x,y)=eaxyH_{\rm ext}+g(y)\ ,
\end{equation}
with $g(y)$ being a function of $y$ only. However, this solution is inconsistent with the second constraint of Eq.~(\ref{scenarioC}). In order to prove this is indeed the case, we can plug the result of Eq.~(\ref{alpha}) into the second equation of (\ref{scenarioC}), which leads to the following condition
\begin{equation}
\partial_{y}g(y)=-e(a+b)xH_{\rm ext}=-exH_{\rm ext}\ .
\end{equation}
which suggests that $g(y)$ must be a function of $x$. This is in contradiction with Eq.~(\ref{alpha}), which suggests that $g(y)$ is a function of $y$ but not $x$. Therefore, scenario C of Eq.~(\ref{scenarios}) cannot be satisfied self-consistently.

\subsection{Scenario B}
Having excluded scenarios A and C, in this subsection we show that scenario B is the only one that can actually be satisfied. Below are two examples:
\subsubsection{Example 1: Normal Vacuum}
A trivial solution that satisfies this constraint is the spatially homogeneous solution with $\pi_{x}=\pi_{y}=\pi_{z}=0$ but $\sigma=f_{\pi}$, which is simply the normal, non-superconducting vacuum of low-energy QCD~\cite{Son1,Son2}. The Gibbs free energy of this state is
\begin{equation}
\label{gibbsnormal}
G_{\rm n}=\int d^{2}\vec{x}\ \frac{1}{2}H^{2}_{\rm ext}\ ,
\end{equation}
with $H_{\rm ext}$ being the external magnetic field.

\subsubsection{Example 2: $\pi^{0}$ domain wall}
Another possible phase that can exist in the Schwinger limit (with no photon back-reaction) is a phase of $\pi^{0}$ domain walls~\cite{Son4}. The phase has the following structure:
In other words, the pion fields are as follows:
\begin{equation}
\begin{split}
\label{dw}
\sigma&=f_{\pi}\cos[4\arctan(\exp(m_{\pi}z)) ]\\
\pi_{z}&=f_{\pi}\sin[4\arctan(\exp(m_{\pi}z)) ]\\
\pi_{x}&=\pi_{y}=0\ .
\end{split}
\end{equation}
\\
\underline{$\mu_{I}=0$}:
This phase was considered in the case of zero-isospin chiral perturbation theory and was calculated by incorporating the effects of the axial anomaly through the Wess-Zumino-Witten term~\cite{Son4, WessZumino, Witten, Kaymakcalan}. Here we briefly summarize the results of Ref.~\cite{Son4}. The Gibbs free energy (not the Gibbs free energy density) of the domain wall phase is
\begin{equation}
\label{gibbsdw}
G_{\rm dw}=\int d^{3}\vec{x}\ \frac{1}{2}H_{\rm ext}^{2}+\int d^{2}\vec{x}\ \left (8 f_{\pi}^{2}m_{\pi}-\frac{eH_{\rm ext}}{2\pi}\mu_{B} \right )\ ,
\end{equation}
where $H_{\rm ext}$ is the external magnetic field and $\mu_{B}$ is the baryon chemical potential.
Furthermore, it was shown in Ref.~\cite{Son4} that the magnetization per unit area of the domain wall state is:
\begin{equation}
\frac{\vec{M}_{\rm dw}}{S}=\hat{z}\frac{e}{2\pi}\mu_{B}\ ,
\end{equation}
and the energy per unit area of the state is 
\begin{equation}
\frac{E_{\rm dw}}{S}=8 f_{\pi}^{2}m_{\pi}\ .
\end{equation}
This phase becomes metastable for external magnetic fields ($H_{\rm ext}$) larger than $\frac{3m_{\pi}^{2}}{e}$, i.e.
\begin{equation}
\label{zeroisospinH}
H_{\rm ext}>\frac{3m_{\pi}^{2}}{e}\ .
\end{equation}
Additionally above a critical magnetic field of $\mu_{B}^{c}$, which is given by
\begin{equation}
\label{criticalmuB}
\mu_{B}^{c}=\frac{16\pi f_{\pi}^{2}m_{\pi}}{eH_{\rm ext}}\ ,
\end{equation}
$\pi^{0}$ domain wall is not only metastable but also becomes stable at the expense of the normal vacuum state if the magnetic is large enough, i.e. satisfies the constraint of Eq~\ref{zeroisospinH}. This can be easily seen by comparing the Gibbs free energies of the $\pi^{0}$ domain wall state in Eq.~(\ref{gibbsdw}) with the Gibbs free energy of the normal state, which is in Eq.~(\ref{gibbsnormal}).\\
\\
\underline{$\mu_{I}\neq 0$}: 
At finite isospin chemical potential, we will show here that while the energy, the critical baryon density and the magnetization of the $\pi^{0}$ domain wall remains unchanged the magnetic fields over which the state becomes metastable changes. Note that it is clear from Eq.~(\ref{Leff}) that the energy is unaffected at finite isospin density as long as charged pions do not condense. Additionally, the magnetization of the domain wall is also unchanged as along as only the charged pions condense.

In particular, the domain wall is stable if
\begin{equation}
\label{finiteisospinH}
H_{\rm ext}>H_{c3}\ ,
\end{equation}
where $H_{c3}$ is defined as follows:
\begin{equation}
\label{Hc3}
H_{c3}=\frac{3m_{\pi}^{2}+\mu_{I}^{2}}{e}\ .
\end{equation}
Note that $H_{c3}$ reduces to the lower bound for the external field of Eq.~(\ref{zeroisospinH}) provided $\mu_{I}=0$.

In order to derive that stability condition, we proceed by expanding the Langrangian of Eq.~(\ref{Lagrangian}) around the $\pi^{0}$ domain wall solution in the $\pi_{\pm}$ directions, which have the following standard definitions:
\begin{equation}
\pi_{\pm}=\frac{\pi_{x}\pm\pi_{y}}{\sqrt{2}} \ .
\end{equation}
Then for small fluctuations of $\pi_{+}$ and $\pi_{-}$, the effective Lagrangian becomes:
\begin{equation}
\label{Lfluctuation}
\begin{split}
\mathcal{L}&\approx(\partial^{\mu}+ieA^{\mu})\pi_{+}(\partial_{\mu}-ieA_{\mu})\pi_{-}\\
&-m_{\pi}^{2}\left (1-\frac{6}{\cosh^{2}(m_{\pi}z)} \right )\pi_{+}\pi_{-}+\mu_{\rm I}^{2}\pi_{+}\pi_{-}\ .
\end{split}
\end{equation}
with corrections of $\mathcal{O}(\pi_{+}^{2}\pi_{-}^{2})$.
%\begin{equation}
%\begin{split}
%\mathcal{L}&=\frac{1}{2}\ [(\partial_{\mu}\pi_{x})^{2}+(\partial_{\mu}\pi_{y})^{2} \ ]\\&-\frac{m_{\pi}^{2}}{2}\left ( 1-\frac{6}{\cosh^{2}(m_{\pi}z)}\right)(\pi_{x}^{2}+\pi_{y}^{2})\\&+\frac{\mu_{\rm I}^{2}}{2}(\pi_{x}^{2}+\pi_{y}^{2})\ ,
%\end{split}
%\end{equation}
Note that this Lagrangian is most easily derived using the parametrization of Ref.~\cite{Son4}. The resulting equation of motion for $\pi_{\pm}$ is as follows:
\begin{equation}
\label{EoMfluctuation}
\begin{split}
-(\partial_{i}\pm ieA_{i})^{2}\tilde{\pi}_{\pm}&+m_{\pi}^{2}\left (1-\frac{6}{\cosh^{2}(m_{\pi}z)} \right )\tilde{\pi}_{\pm}\\&-\mu_{\rm I}^{2}\tilde{\pi}_{\pm}=\omega_{n}^{2}\tilde{\pi}_{\pm}\ ,
\end{split}
\end{equation}
where we have used to decomposition 
\begin{equation}
\pi_{\pm}(\vec{x},t)=\sum_{n}e^{i\omega_{n}t}\tilde{\pi}_{\pm}(\vec{x})\ .
\end{equation}
The potential of Eqs.~(\ref{Lfluctuation}) and (\ref{EoMfluctuation}) in the absence of an external magnetic field is a standard quantum mechanical reflection-less potential. In a magnetic field the dispersion relation is simply:
\begin{equation}
\omega_{n}^{2}=(2n+1)qH-3m_{\pi}^{2}-\mu_{I}^{2}\ .
\end{equation}
From the dispersion relation, it is obvious then that the $\pi^{0}$ domain walls are metastable only if $\omega_{n}^{2}$ is positive for all $n$. It is easy to see then that this occurs if the condition of Eq.~(\ref{finiteisospinH}) is satisfied. While the metastability condition has changed going to the finite isospin case, note that the energy per unit area of the $\pi^{0}$ domain wall phase remains unchanged since the isospin chemical potential enters the Lagrangian through a term proportional to $\pi_{+}\pi_{-}$, which does not contribute to the energy. Furthermore, these equations of motion remain valid even with the inclusion of the Wess-Zumino-Witten term in chiral perturbation theory since the contribution to the action of this term arises from a full derivative term, which does not affect the equations of motion.

\begin{figure}[h]
    \centering
    \includegraphics[width=0.5\textwidth]{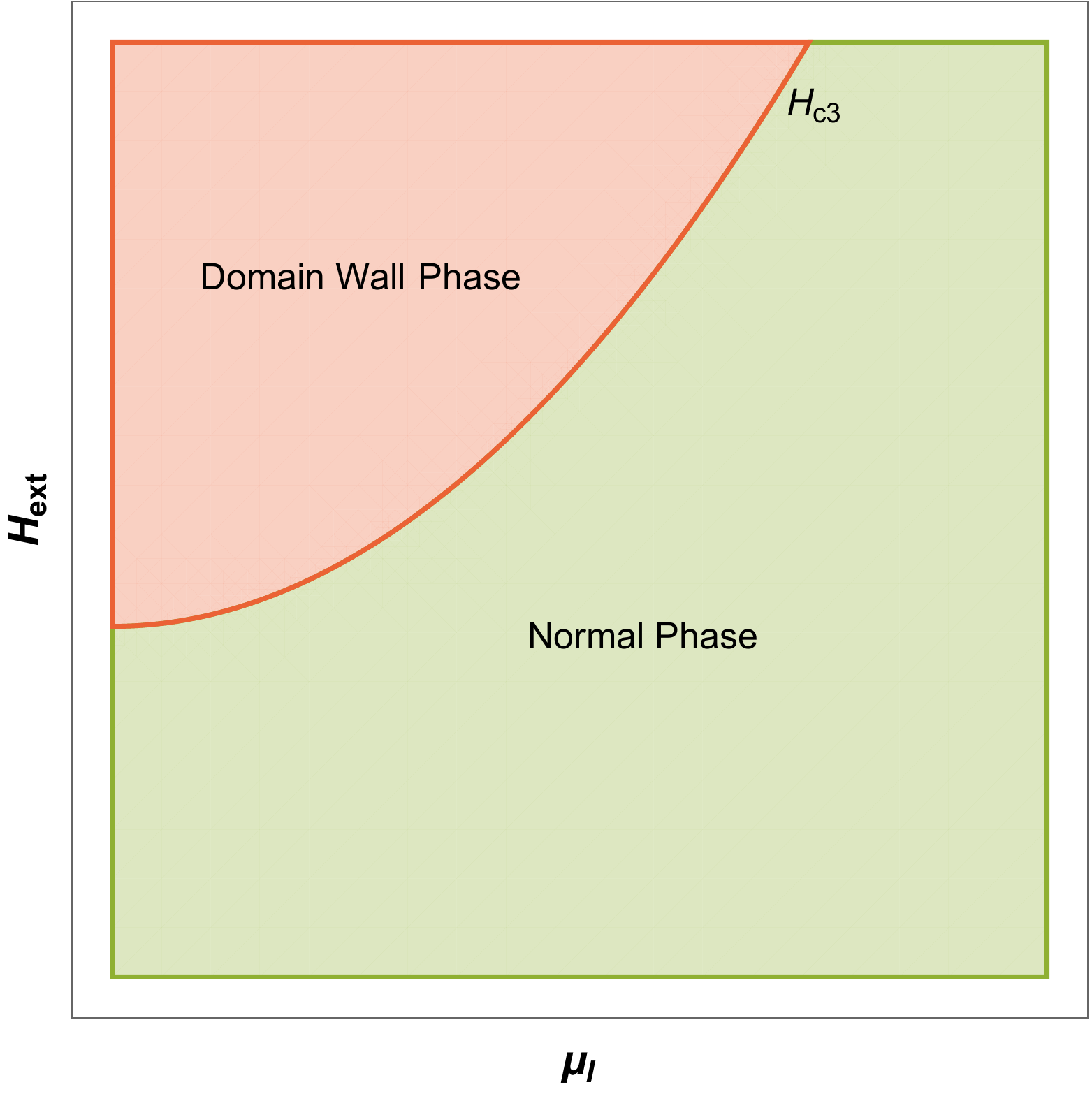}
    \caption{A schematic plot showing the two phases that can exist above the critical baryon chemical potential in the Schwinger limit. The isospin chemical potential is plotted on the x-axis while the external magnetic field $H$ is plotted on the y-axis. $H_{c3}$ is the phase transition line from the normal phase to the domain wall phase and it depends on the isospin chemical potential through Eq.~(\ref{Hc3}).}
    \label{belowmuB}
\end{figure}

\section{Phase Diagram at finite isospin density and finite baryon density}
\begin{figure}[h]
    \centering
    \includegraphics[width=0.5\textwidth]{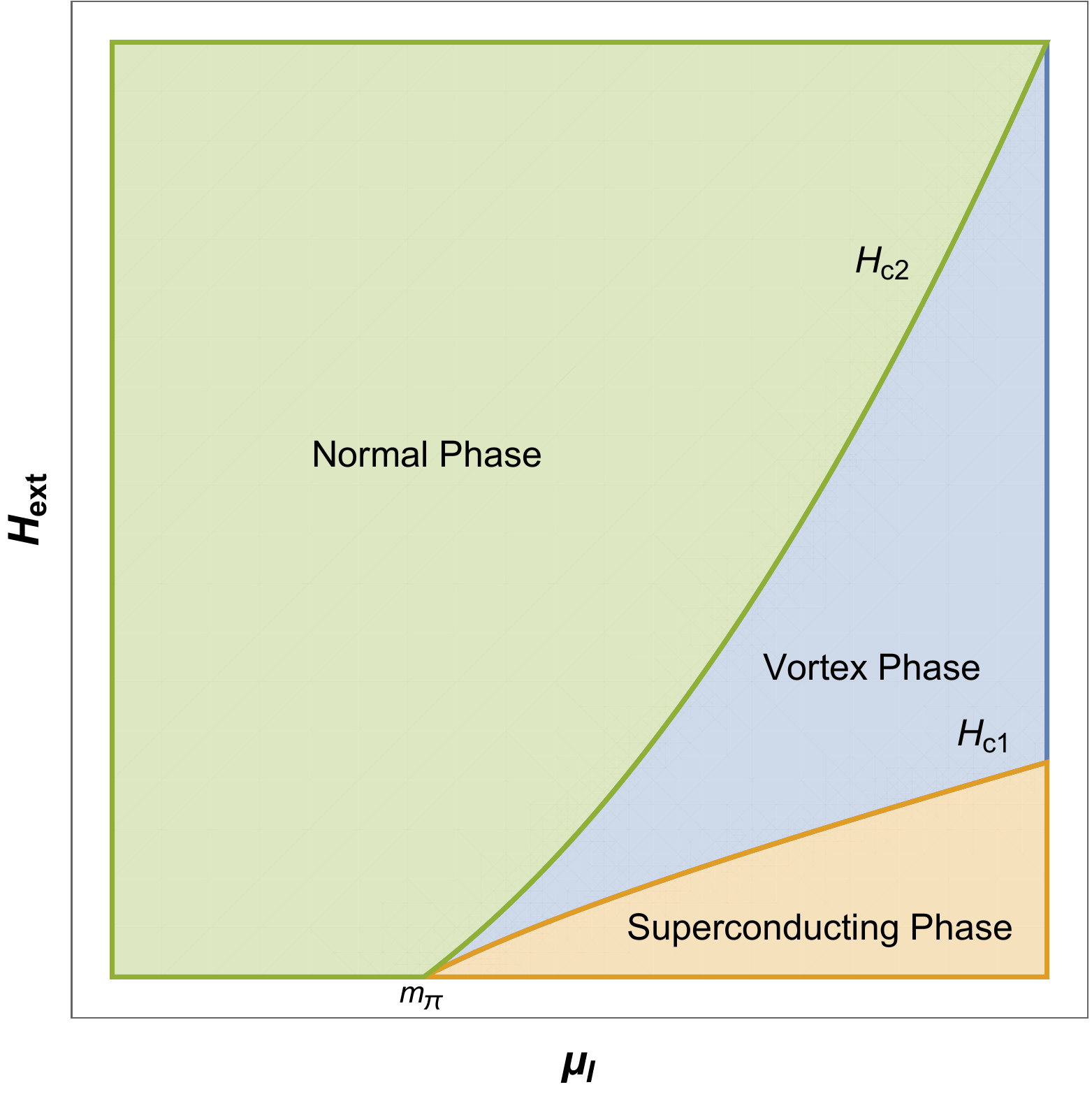}
    \caption{A schematic plot showing the different phases that exist below the critical baryon chemical potential. Note that for isospin chemical potentials, $\mu_{I}$ below $m_{\pi}$, the normal vacuum state persists. Above $m_{\pi}$ and below $H_{c1}$, a uniform superconducting state exists; between $H_{c1}$ and $H_{c2}$, there is an inhomogeneous phase of superconducting vortices and above $H_{c1}$ the normal state is energetically favored. (color online)}
    \label{belowmuB}
\end{figure}
\subsection{Schwinger Limit: No Photon Back-reaction}
Now we can consider the phase diagram of finite isospin chiral perturbation theory in the Schwinger limit but in the presence of a baryon chemical potential, which we will assume is smaller than the mass of the lightest nucleon such that nuclear matter does not condense. Note that if there is no baryon chemical potential the normal phase is energetically favored and there are no phase transitions. The only other phase we know of that can occur in the Schwinger limit is the $\pi^{0}$ domain wall state, which only becomes stable above the critical baryon chemical potential of Eq.~(\ref{criticalmuB}).

\begin{figure}[h]
    \centering
    \includegraphics[width=0.5\textwidth]{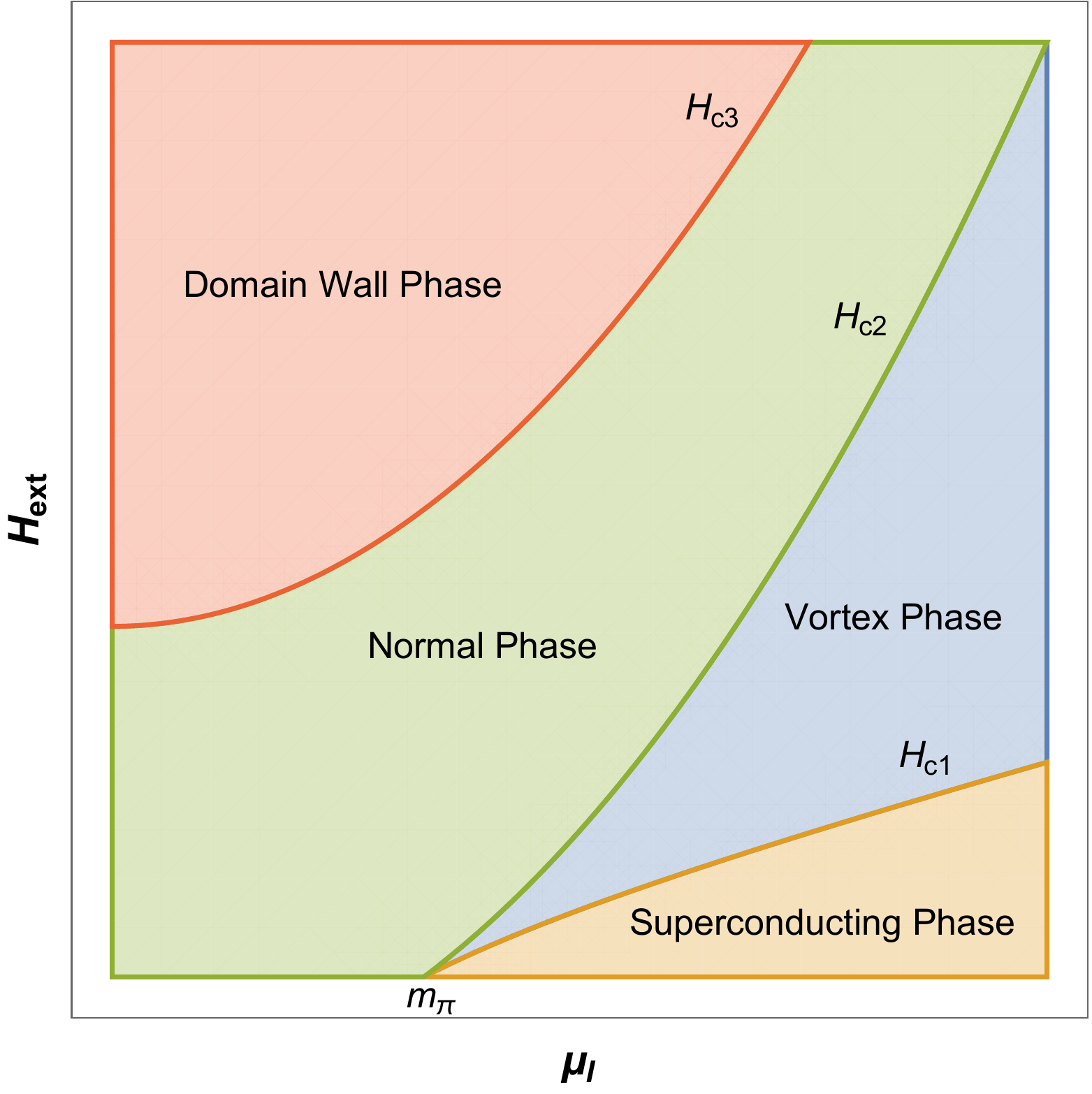}
    \caption{A schematic plot showing the different phases that exist above the critical baryon chemical potential. Note that for isospin chemical potentials, $\mu_{I}$ above $m_{\pi}$ and below $H_{c1}$, a uniform superconducting state exists; between $H_{c1}$ and $H_{c2}$, there is an inhomogeneous phase of superconducting vortices and above $H_{c1}$ the normal state is energetically favored. Above $H_{c3}$, even if $\mu_{I}<m_{\pi}$, the domain wall phase is more stable than the normal state. (color online)}
    \label{above-muB}
\end{figure}

\subsection{Case with photon back-reaction}
The $\pi^{0}$ domain wall state remains stable away from the Schwinger limit but is energetically disfavored compared to the normal vacuum, the vortex state and the uniform superconducting state of finite isospin chiral perturbation theory when $\mu_{B}=0$. However, with the inclusion of anomaly effects through the WZW term, $\pi^{0}$ domain walls do become stable above the critical baryon chemical potential defined in Eq.~(\ref{criticalmuB}).

In Fig~\ref{belowmuB}, we show the phase diagram of finite isospin chiral perturbation theory in the absence of a baryon chemical potential. This phase diagram was considered in Ref.~\cite{Adhikari}, where we assumed that while the external magnetic field couples to the charged pions, the pions themselves only interact with each other through strong interactions at leading order in chiral perturbation theory. This is analogous to the nuclear matter problem, where electromagnetic effects are ignored, and the physical picture is valid as long $\mu_{I}\gg\mu_{\rm es}$, where $\mu_{I}$ is the finite isospin chemical potential, the amount of energy for an additional charged pion to condense and $\mu_{\rm es}$ is the electromagnetic energy associated with adding the additional charged pion.

Above the critical baryon chemical potential of Eq.~(\ref{criticalmuB}) and below nuclear saturation densities, i.e. the silver blaze region for the baryon chemical potential, the $\pi^{0}$ domain wall state becomes energetically more stable compared to the normal phase. In order to characterize the phase diagram of finite isospin chiral perturbation theory in a magnetic field ($H_{\rm ext}$) above the critical baryon chemical potential, it is important to determine exactly the critical magnetic field, $H_{c2}$, which is the external magnetic field where the transition from the vortex state to the normal state occurs. Close to this critical point, the magnitude of the charged pion condensates are small compared to the pion decay constant, i.e. $\theta\approx 0$. As such the Lagrangian is equal to
\begin{equation}
\begin{split}
\mathcal{L}_{\rm eff}=&\frac{1}{2}H_{\rm ext}^{2}-\frac{1}{2}\left (\partial_{i}+i e A^{\rm ext}_{i} \right ) \pi_{+}\left ( \partial_{i}-i e A^{\rm ext}_{i} \right )\pi_{-}\\
&-\frac{1}{2}\left (\mu_{I}^{2}-m_{\pi}^{2}\right ) \pi_{+}\pi_{-}+\mathcal{O}\left ( (\pi_{+}\pi_{-})^{2} \right)\ ,
\end{split}
\end{equation}
and has corrections that are quartic in the charged pion fields. Note that $A^{\rm ext}$ is the vector potential associated with an approximately uniform magnetic field of strength $H_{c2}$, i.e. the second critical point, pointing in the z-direction. In writing the Lagrangian down, we assumed that the neutral pion field remains uncondensed. It was shown numerically in the low density regime of Abrikosov vortices that the neutral pion does not condense. We assume that this remains the case near the second critical field, where the vortices are densely packed. The equation of motion for the charged pions is then
\begin{equation}
-(\partial_{i}-ie A_{i}^{\rm ext})^{2}\pi_{+}+m_{\pi}^{2}\pi_{+}=\mu_{I}^{2}\pi_{+}\ ,
\end{equation}
and the dispersion relation of this gives the standard Landau levels
\begin{equation}
\mu_{I}^{2}=(2n+1)eH_{\rm ext}+k_{z}^{2}+m_{\pi}^{2}\ .
\end{equation}
Given this dispersion relation, the largest magnetic field that can be sustained in an Abrikosov vortex lattice occurs when $n=0$ and $k_{z}=0$. Therefore, the largest value that can be assumed by $H_{\rm ext}$ is the critical field $H_{c2}$, which is as follows:
\begin{equation}
eH_{c2}=\mu_{I}^{2}-m_{\pi}^{2}\ .
\end{equation}

Now we know the two critical fields, $H_{c2}$ and $H_{c3}$, where the transition from the vortex state to the normal state and the transition from the normal state to the $\pi^{0}$ domain wall state occurs respectively, we can proceed to construct the phase diagram of finite isospin chiral perturbation theory in a uniform magnetic above the critical magnetic field and the critical baryon density. The phase diagram is present in Fig~\ref{above-muB}. Note that the difference between $H_{c3}$ and $H_{c2}$ is
\begin{equation}
H_{c3}-H_{c2}=\frac{4m_{\pi}^{2}}{e}\ .
\end{equation}
However, in the chiral limit, the phase transition to the $\pi^{0}$ domain wall state occurs directly from the vortex state without first a phase transition to the normal state that occurs for finite pion masses. Note that in the chiral limit, $H_{c3}=H_{c2}$, suggesting that the normal phase is absent in the phase diagram.

\begin{acknowledgements}
I would like to acknowlege the support of the physics department at St. Olaf College.
\end{acknowledgements}

\end{document}